# Clusters of microparticles in distilled water: a kaleidoscope of versions and paradoxes of nature (Review)

Tatyana Yakhno [1]*, Vladimir Yakhno [1,2]

[1] Federal Research Center Institute of Applied Physics of the Russian Academy of Sciences (IAP RAS), 46 Ulyanov Street, Nizhny Novgorod 603950, Russia. yakhta13@gmail.com

[2] N. I. Lobachevsky State University of Nizhny Novgorod (National Research University), Nizhny Novgorod 603950, Russia. yakhno@ipfran.ru

\* Correspondence: yakhta13@gmail.com (TY); yakhno@appl.sci-nnov.ru (VY);

Tel.: (011)-7-831-436-85-80 (TY and VY)

Abstract

The presence of microparticles (clusters of micron size) of unknown origin in the volume of water, including highly purified water (bidistilled, deionized), has been repeatedly demonstrated by various methods of physical analysis. Various assumptions have been made about the nature of these microparticles, but none of them has become generally accepted. The review analyzes the literature data and the results obtained by the authors using optical and electron scanning microscopes. The composition and phase state of distilled water deposits at the bottom of glassware after evaporation of free water are considered. The structure of the microdispersed phase of distilled water and the mechanism of phase transitions of its components in the process of natural evaporation, the end products of which are gel-like water and sodium chloride crystals, are proposed.

**Keywords:** distilled water; dispersed phase; hydrated water; evaporation; phase transitions; sodium chloride.

Introduction

    1.1. Current state of the problem

The structure of water is traditionally considered on the scale of nanometers [1–6]. However, to date, using various physical research methods, evidence has been obtained for the structural inhomogeneity of water on scales from units to hundreds of micrometers [7–26]. Researchers are unanimous that high-purity (high-resistance) water contains molecular associates - clusters ranging in size from 2.3 to 120 microns. But opinions about their nature and formation mechanisms depend on the instrumental technique used and the area of special competence of the authors (Table 1).

We observed the structure of water samples using a conventional optical microscope in a "crushed drop" preparation with a layer thickness of ~8 μm [23, 24]. The experiments were carried out under laboratory conditions at T = 22–24°C, H = 73–75%. For microscopic observations, we used distilled water (TU 2384-009-48326337-2015, specific electrical conductivity 4.5 μS/cm, pH 7.0), tap water (specific electrical conductivity 550 μS/cm), and extra pure water (OST 34-70 -953.2-88, electrical conductivity 0.04–0.05 μS/cm, pH 5.4–7.0).

# Studies of microstructural inhomogeneities in the volume of water

Tab. 1.

| № | Authors | Particle size (μm) | Instrument | Suggested nature of the particles |
|---|---------|--------------------|-----------|-----------------------------------|
| 1. | N.F. Bunkin et al. [7-9] | Characteristic particle radius ≅ 0.5 μm and fractal dimension within 2.5–2.8 | Laser modulation-interference phase microscopy and laser light scattering | Clusters of air nanospheres (bubstons) |
| 2. | A.N. Smirnov et al. [10,11] A.V. Syroeshkin [12] V.V. Goncharuk et al. [13] | Giant Heterophasic Clusters (GHC) of water - long-lived inhomogeneities with relaxation times of more than 10 seconds. 5 populations of clusters ranging in size from 0.5 to 120 μm were identified | Laser Small Angle Dispersion Meter (particle sizer); NMR-(spin-spin relaxation of protons); non-inertial electrothermometer | The deuterium concentration determines the structural features of water, in particular, the size of GHCs and their stability |
| 3. | A.N. Smirnov [14] | Supramolecular water complexes - "emulons" 1-100 microns (5 fractions of clusters - from 1 to 120 microns). Relaxation time - over 1 sec. | Acoustic emission, laser light scattering, thermometric analysis | Under the influence of hydrated hydrogen and hydroxyl ions, the system self-organizes with the formation of spatial structures (emulons). |
| 4. | B.I. Laptev et al. [15] | In distilled water, there are associates up to $10^{-6}$ M in size (1 μm). When NaCl is added, the mobility of water dipoles in them, in comparison with distilled water, increases in proportion to the concentration. | Dielectrometry and resonance method | The decrease in electrical capacitance with increasing current frequency is due to the existence of molecular associates (clusters) in water. The increase in the capacity of salt solutions is due to the processes of deassociation of water clusters and hydration of ions. |
| 5. | V.I. Bukaty, P.I. Nesteryuk [16,17] | The size spectrum of optical inhomogeneities (clusters) was 1.5–6.0 μm, while the arithmetic mean radius was 2.3 μm, and the root-mean-square radius - 2.5 μm. | Measuring and computing complex using the method of small scattering angles | Not considered |
| 6. | M.J. Sedlak [18-21] | Submicron-sized domains (large clusters) with higher solute concentration than in the rest of solution. | Static and dynamic laser light scattering | Obtained results show that these structures are not nanobubbles in all cases. |
| 7. | E.E. Fesenko, E.L. Terpugov [22] | 10 – 1000 μm | IR spectroscopy of a thin layer of water | Not considered |
| 8. | T. Yakhno, V. Yakhno [23,24] T.Yakhno, M.Drozdov, V. Yakhno [25] | 10 – 1000 μm | Optical microscopy of a thin layer of water | NaCl microcrystals surrounded by a liquid crystal hydration shell |
| 9. | T. Yakhno et al. [26] | 5 – 10 μm | Scanning electron microscope | NaCl microcrystals surrounded by a liquid crystal hydration shell |

We made sure that the structure of water of different nature is qualitatively the same (Fig. 1,2). This fact suggests that the presence of microstructures in water is its inherent property.

The study of a thin layer of distilled water (~8 μm) through an optical microscope made it possible to detect the presence of microparticles of the dispersed phase (DP) in it, i.e., transparent spherical elements ~10 μm in diameter with a dark particle in the center [23]. The microparticles combined into aggregates up to several hundred micrometers in size. As our studies have shown, the central structure-forming particle of DP is a sodium chloride microcrystal, and the sphere surrounding it is a liquid-crystal shell of hydration water. The composition, physical properties, and possible origin of DP water are discussed in [23, 24]. The dynamics of phase transitions during the drying of thin water films with the formation of large sodium chloride crystals is also considered there. The study of the physical properties of DP showed, in particular, that its aggregates are similar to a viscous liquid, do not evaporate at room temperature, do not dissolve in organic solvents, are eroded in salt solutions, and evaporate at a temperature of about 300°C [26].

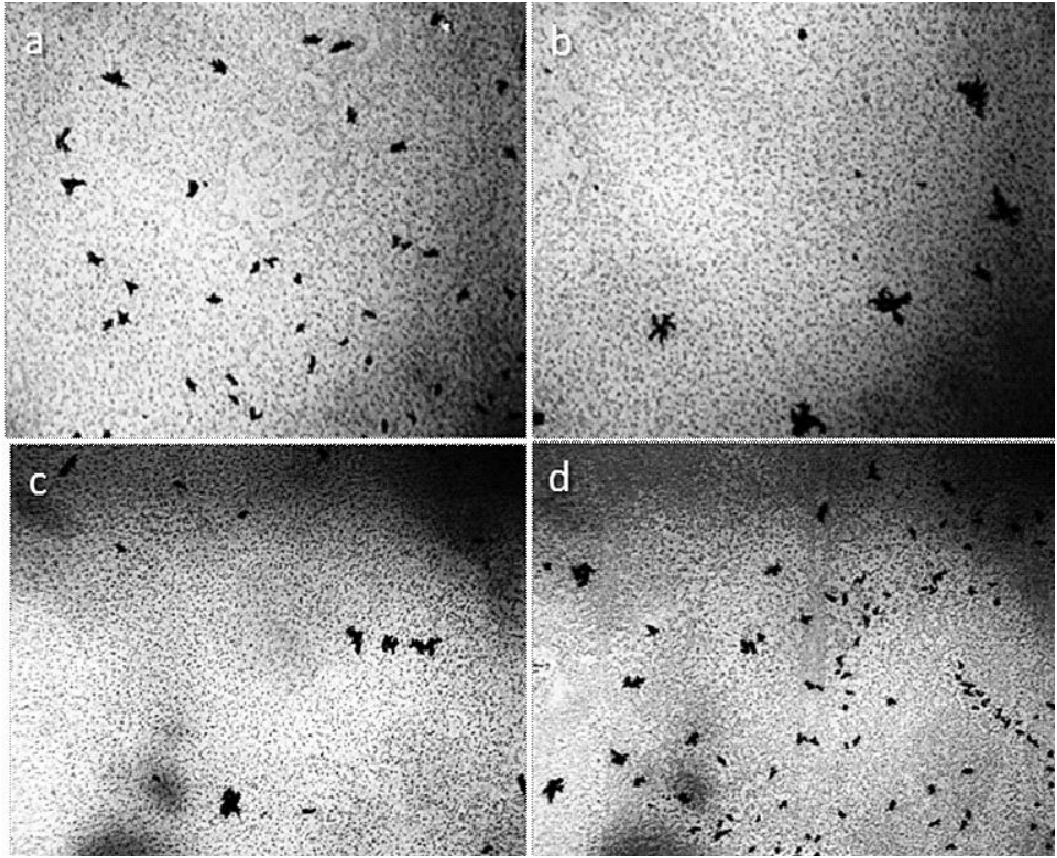

Figure 1. The structure of "ultrapure" water (a,b) immediately after depressurization the container essentially corresponds to the structure of distilled (c) and mineral (d) water. The width of each frame is 3 mm. The liquid layer thickness is 8 μm [26].

Using the same microscope, we observed the structure of water placed in a through hole in a plastic plate. This made it possible to consider a drop of liquid "through" (Fig. 2). The diagram shows the relative solids content of each sample after the free water has evaporated.

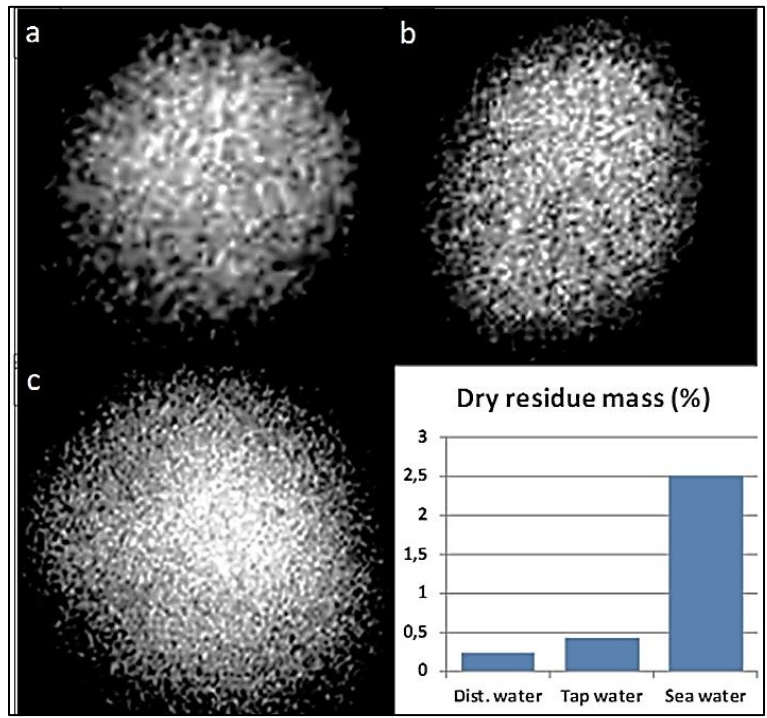

Figure 2. Microstructure of liquids in the "hanging drop" preparation. Liquid placed in a hole in a plastic plate with a diameter of 0.5 mm and 0.5 mm thickness. Microscopy "through liquid" – a: distilled water; b: tap water; c: water of the Black Sea [26]. The diagram shows the relative solids content of each sample after the free water has evaporated.

A team of authors from a number of well-known Russian scientific institutions published the results of their studies of deionized water using a Malvern laser small-angle dispersity meter [12].

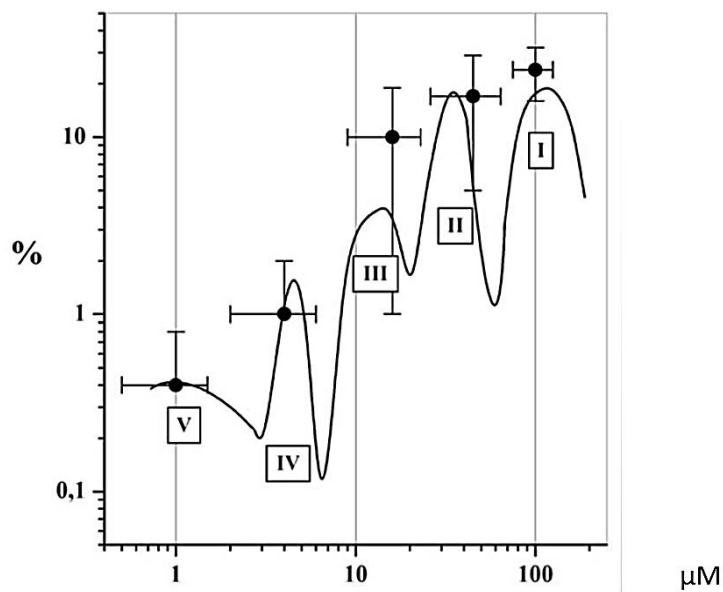

Figure 3. Size spectra of Giant Heterophase Water Clusters. Potentially possible populations (based on the analysis of 20 preparations of deionized water obtained at different times) [12].

The size distribution of particles indicates the possibility of their observation under an optical microscope. We compared our results with the results of other authors, noting similarities and differences (Table 1).

The sizes of optical inhomogeneities found by different authors are mainly from a few to 1000 μm. In our opinion, this is due to different sizes of DP aggregates [26]. We have previously shown that DP is prone to aggregation. The aggregates reversibly decomposed under the energy impact on water of various physical agents: - heat, low-intensity laser radiation, mechanical mixing, ultrasound. The disintegration of aggregates (confirmed microscopically) was invariably accompanied by regular changes in the physicochemical properties of the solution: an increase in pH, electrical conductivity and sound speed and a decrease in electrical capacitance, viscosity, surface tension and redox potential. These changes were associated with an increase in the area of the interfacial surface [26].

According to the authors of [10–14], the Giant Heterophasic Clusters (GHCs) registered by them are dynamic water associates with a relaxation time from fractions to units of a second. According to our data, the observed DP aggregates are quite stable, subject to Brownian motion in the volume of water, and disintegrate only under the influence of a number of agents of a physical nature (see above).

We consider the term "emulons" [15] to be incorrect, since each structure, in addition to the liquid of another phase, also contains a particle of a solid substance - a microcrystal of sodium chloride. In this case, any transitional forms were not observed. A number of authors insist that the observed microstructures consist of gas microbubbles (bubstons) [7–9]. In the works of other authors, this idea has not been confirmed [18-21; 23,24].

The noted contradictions do not in the least discredit the authors of the studies, since they are dealing with a complex open system that is in a stably nonequilibrium state [27] and consider this system based on their past experience, each in his own way. We would like to present here some additional information that would allow a broader view of the problem in its entirety.

1. Where does salt (NaCl) come from in microparticles?

This question has been of interest to us since we discovered the presence of sodium chloride (confirmed crystallographically) at the bottom of a glass dish after a volume of distilled water has evaporated from it [23]. Mindful of the unity of the world around us, we assumed that salt got into the water as air pollution. Salt in the form of microcrystals is also present in freshly melted snow (Fig. 4) [26].

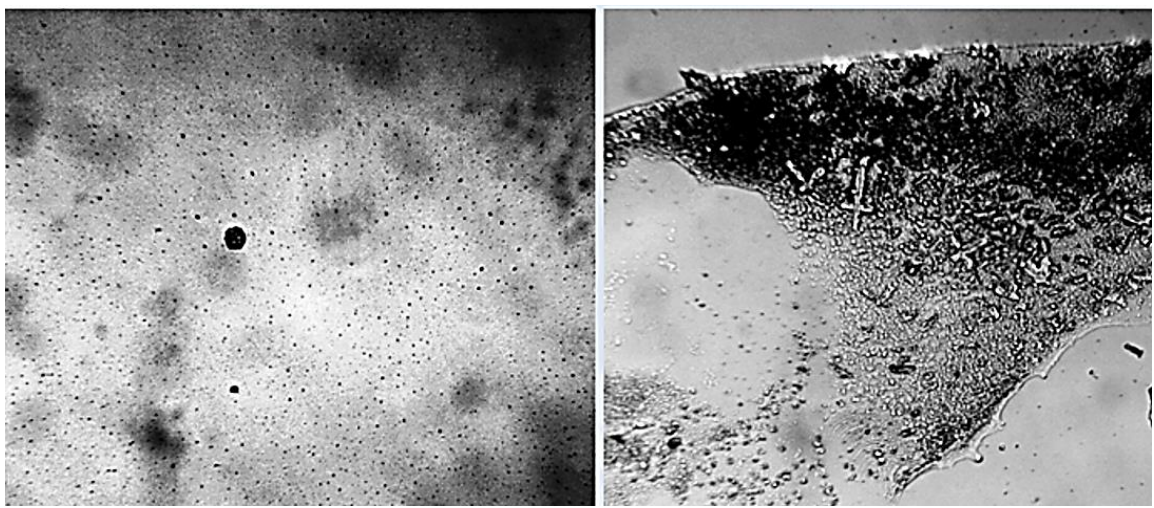

Figure 4. Left - the structure of freshly melted snow. The arrangement of salt microcrystals resembles a "colloidal crystal". On the right is a non-drying film of water with salt crystals 7 days after the evaporation of free water. The width of each frame is 3 mm [26].

To test the possibility of salt ingress from the ambient air into water, the following experiment was carried out [24]. 50 ml of distilled water from the same container was poured into two identical clean glass beakers. The end of a plastic tube connected to an aquarium compressor was immersed into one of the beakers. The input end of the tube was placed above the laboratory table in the working area. Laboratory air was passed through the water for 10 min, at a rate of 72 l / h (before the experiment, the compressor was idling for 20 min to clean air paths from possible internal contaminants). Thus, 12 l of air passed through 50 ml of water during the experiment. Both beakers (control and experimental) were left on the table at room conditions, covered with a flat lid, for a week to restore the structural balance. A week later the electrical conductivity of the control sample was 4.1 µS/cm, and of the experimental one 5.4 µS/cm. The microscopy of the droplets of the water samples tested on the glass slide also revealed the difference in the content and structure of the sediment (Fig. 5).

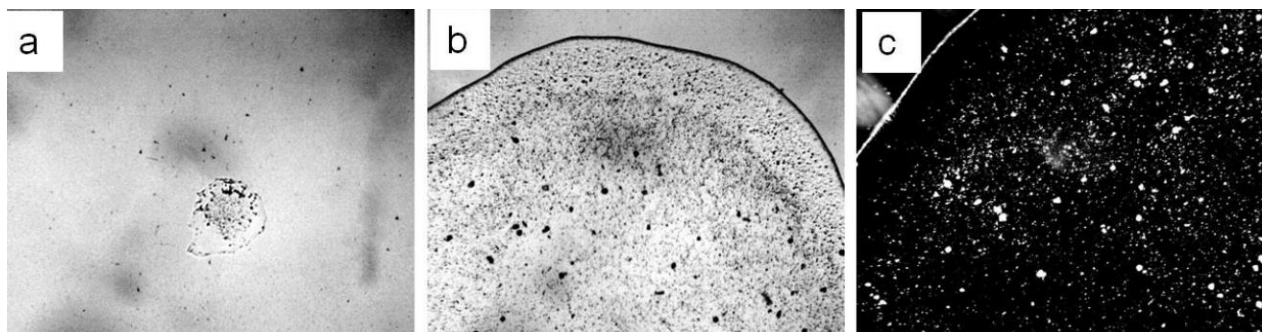

Figure 5. Microphoto of water droplets dried on glass: a - control; b, c - after passing the air (c - dark-field image). Frame width: a, b - 3 mm, c - 1 mm [24].

According to the report of the Intergovernmental Panel on Climate Change (IPCC) for 2001, the annual release of sea salt (NaCl with an admixture of K+, Mg2+, Ca2+, SO42−) from the ocean surface into the atmosphere is 3300 megatons per year [28]. The size of the salt crystals in the atmosphere is, in general, a few microns or less with a predominance of micrometer particles. According to [29], the size of salt crystals in the atmosphere can reach 100 microns. A significant part of NaCl also enters the atmosphere as part of industrial emissions, volcanic activity, vehicular pollution, and human economic activity. The size of the salt crystals in the atmosphere is mostly from fractions to a few microns with a predominance of micrometer particles. These particles begin their existence as droplets of seawater with diameters of 0.1-100 µm. Depending on the temperature and vapor pressure of water, these droplets may become more or less concentrated, or they may crystallize [30]. Laboratory experiments have shown, that NaC1 particles below 44% RH can be expected to be crystalline regardless of how they were generated, and their water content must be explained in terms of entrainment in the particles or adsorption on the surfaces of the particles [30]. Generalizing the literature data, we can assume that if the microcrystalline NaCl coated with the hydration shell, formed in the atmosphere, falls into liquid water, then it has every chance to maintain its integrity, because the thick hydrated shell will protect it from dissolution. However, most of the sea salt that has not been in close contact with the air remains in solution. Although a number of authors, on the basis of molecular modeling, believe that NaCl crystallization can also occur in aqueous solutions [31]. It should be noted that both a deficiency and an excess of sodium chloride in the human body leads to serious disturbances in the water balance, up to death [32].

Thus, we assumed that salt microparticles surrounded by a thick layer of hydrated liquid crystal water (Exclusion Zone [33]) [23] can enter the water from the air. After complete evaporation, large NaCl crystals and gel-like non-evaporating water remain on the glass substrate (Fig. 6 [25]). The use of a JEOL JSM – 6390 LA scanning electron microscope made it possible to reveal a number of previously unknown details: the formation of small crystals of sodium chloride on the surface of the dispersed phase of distilled water (Fig. 7) [25] and the growth of crystals on strands of gel-like water (Fig. 8) [25 ].

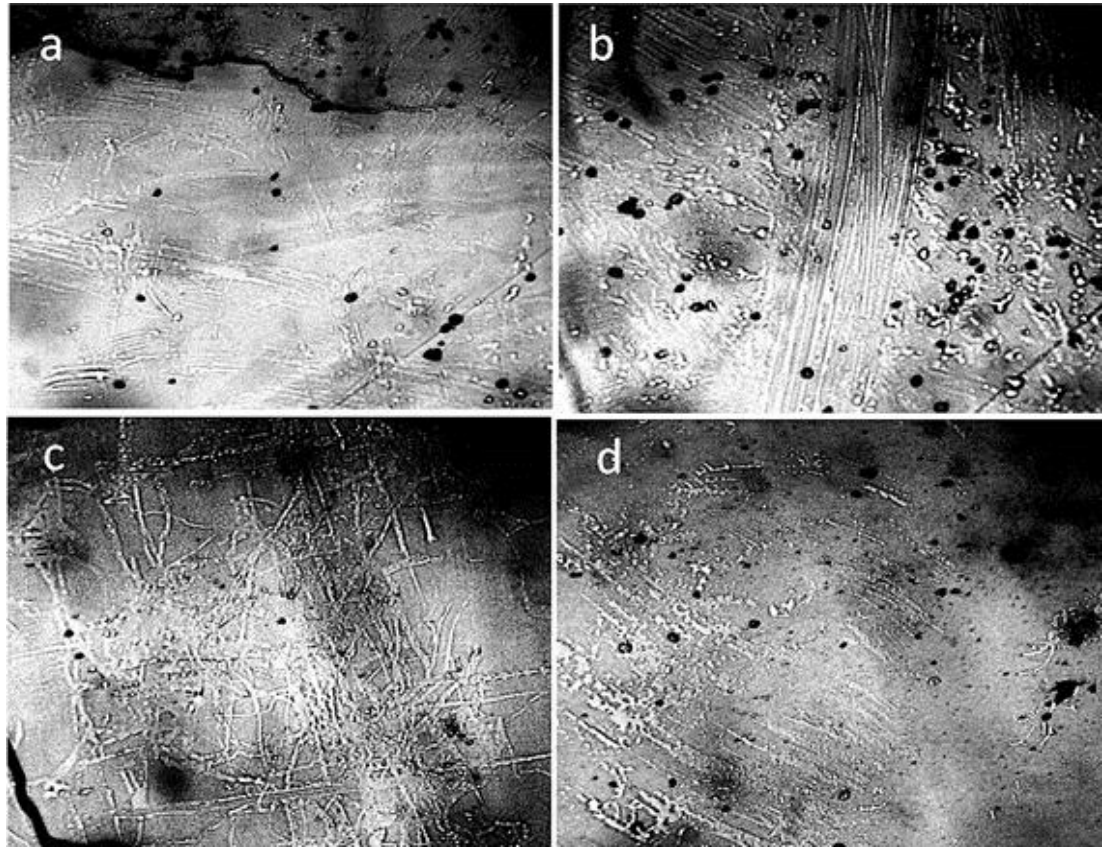

Figure 6. Precipitation on glass after evaporation of free water from a glass surface under an optical microscope (different fields of view). Strands of non-evaporating gel-like water and NaCl crystals are visible (dark dots). Figure (b) shows the mark of a wooden toothpick that deformed the gel-like deposits [25]. The width of each frame is 3 mm.

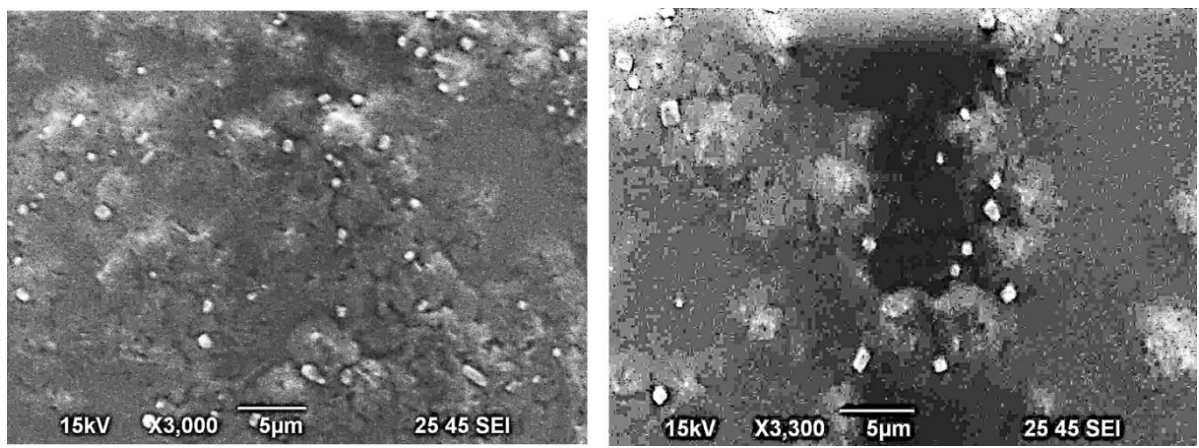

Figure 7. Scanning electron microscope. The dispersed phase of distilled water at the bottom of a glassware after the free water has evaporated. Small NaCl crystals are visible above the stroma of DP [25].

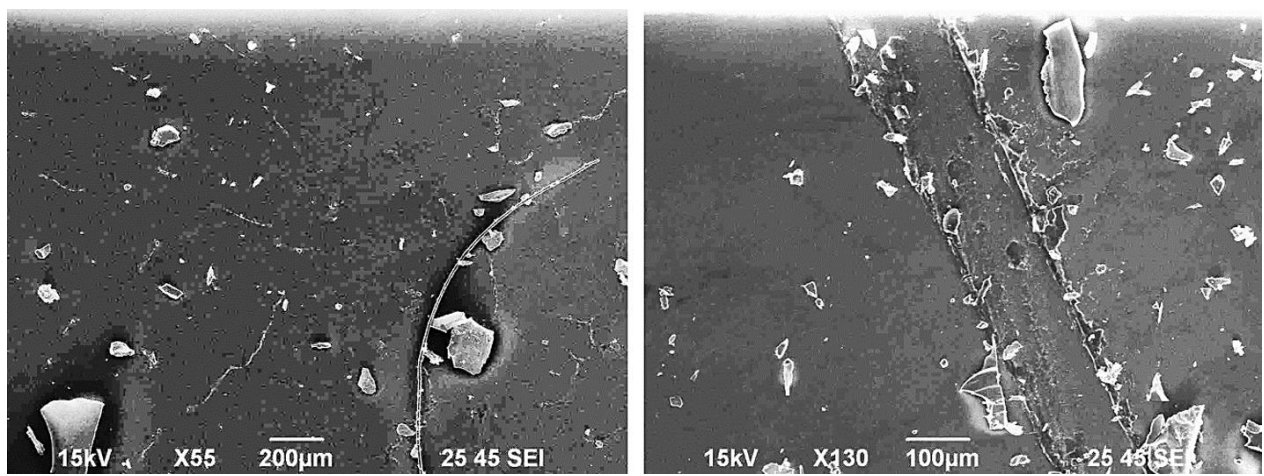

Figure 8. Scanning electron microscope. Gel-like strands of non-evaporating water with NaCl crystals germinating from them [25].

Based on our observations [24] and information about the structure and properties of the "Exclusion Zone" (EZ) [33, 34], we presented a scheme of phase transitions of water components during the evaporation of free water (Fig. 9).

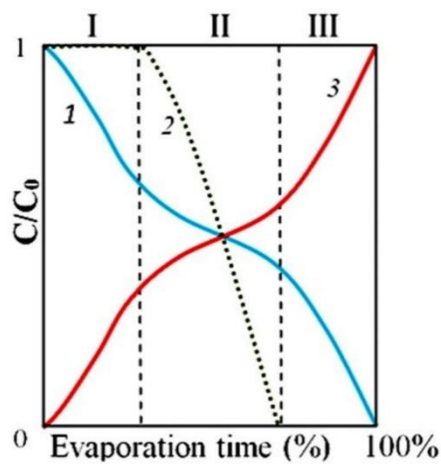

Figure 9. Schematic diagram of the dynamics of phase transitions in water containing Liquid Crystal Spheres with NaCl microcrystals as a "seed" during its evaporation from a solid substrate in the "Relative concentration (C/C0) – Evaporation time (%)" coordinates. Stage I—evaporation of free water and increase in osmotic pressure; Stage II—phase transition of liquid-crystal water into free water and reducing relative rate of evaporation; Stage III—growth of NaCl crystals. 1(blue) - relative concentration of free water; 2 (dotted line) - relative concentration of liquid-crystal water; 3 (red) - relative salt concentration [24].

Evaporation of free water is accompanied by an increase in osmotic pressure in the remaining liquid volume, which leads to the onset of melting of hydrate shells [23, 24] and partial exposure of salt microcrystals contained inside them. The photographs presented here show that the upper part of the DP elements is covered with small NaCl crystals. In addition, larger crystals can be observed on top of the DP layer. The structure and properties of hydrate shells correspond to the structure and properties of EZ [33–38]. According to the authors of [34], EZ has a spongy structure, where the walls of the sponge are represented by a dense highly structured aqueous phase, and the cells are filled with ordinary water. Like a hydrogel [39], this structure has a high absorbency. As free water evaporates, the water enclosed in hydration shells evaporates more slowly. This fact should be taken into account when considering the dynamics of phase transitions in the system under consideration. As for the strands of non-evaporating gel-like water, when viewed under SEM, it becomes noticeable that they include small crystalline formations that are capable of growing and destroying these structures (Fig. 8). If the strands consist of EZ, then the salt structures growing on them "pull out" the free water contained in them, which leads to a violation of the integrity of these strands. Figure 10 summarizes our understanding of the process.

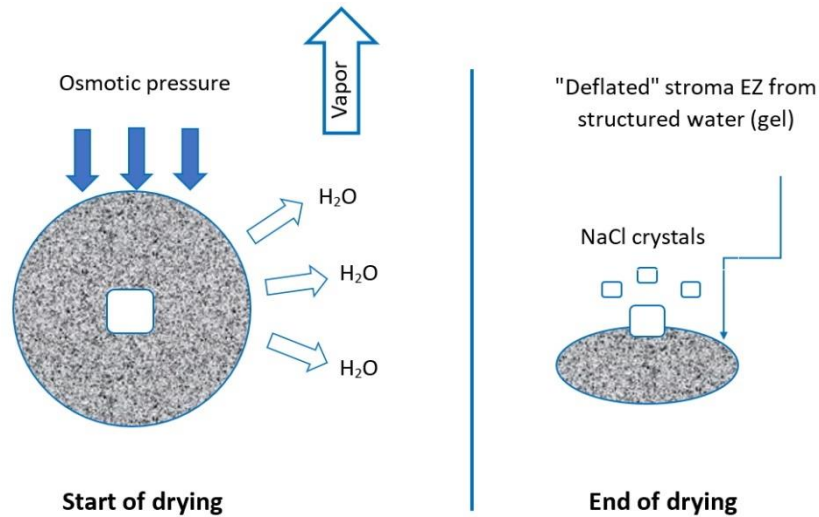

Figure 10. Schematic representation of the processes in the microparticles of the dispersed phase, caused by the evaporation of the surrounding free water.

At this stage of research, we believe that the loose structureless mass in Fig. 7 represents the remnants of the DP stroma. Under an optical microscope, it looks like a gel (Fig. 6).

It is known that EZ is a loose polymer of water molecules formed at hydrophilic surfaces [33, 34, 40]. This polymer structure displaces most of the ions and microparticles from its volume and has a negative charge of the surface in contact with ordinary water of the order of -120 mV. It is assumed that the interaction of microparticles with each other and with near-wall water is carried out according to the "Like likes like" mechanism [42], i.e., through ions of the opposite sign (+) accumulating near EZ, which carries a negative charge [33]. We conducted an experiment in which slides were placed in glasses with distilled and tap water and kept there for two days (Fig. 11 [43]).

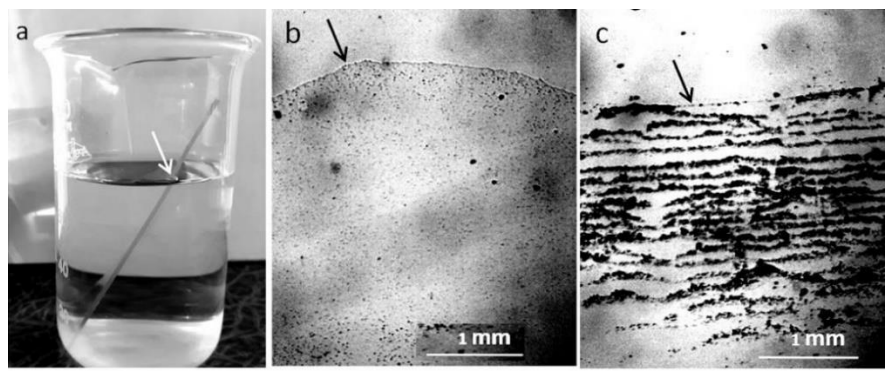

Figure 11. Scheme of the experiment (a) and the structure of deposits on glasses after two days of their incubation in distilled (b) and tap (c) water (Microscopic Photograph). The arrows indicate the initial boundary between the glass and the water surface [43].

As can be seen in the photographs, the deposits on the glass slide (Figure 11b) after its contact with distilled water are an even layer of densely packed rounded structures with a dark dot in the center, which we previously called the dispersed phase of water (DP). The upper part of the glass incubated in tap water (Figure 11c) is covered with bands of deposits of coarse microimpurities, reflecting the decreasing of the level of evaporating water. The deposition of DP aggregates in the middle part of the glasses can be multilayered (Figure 12d).

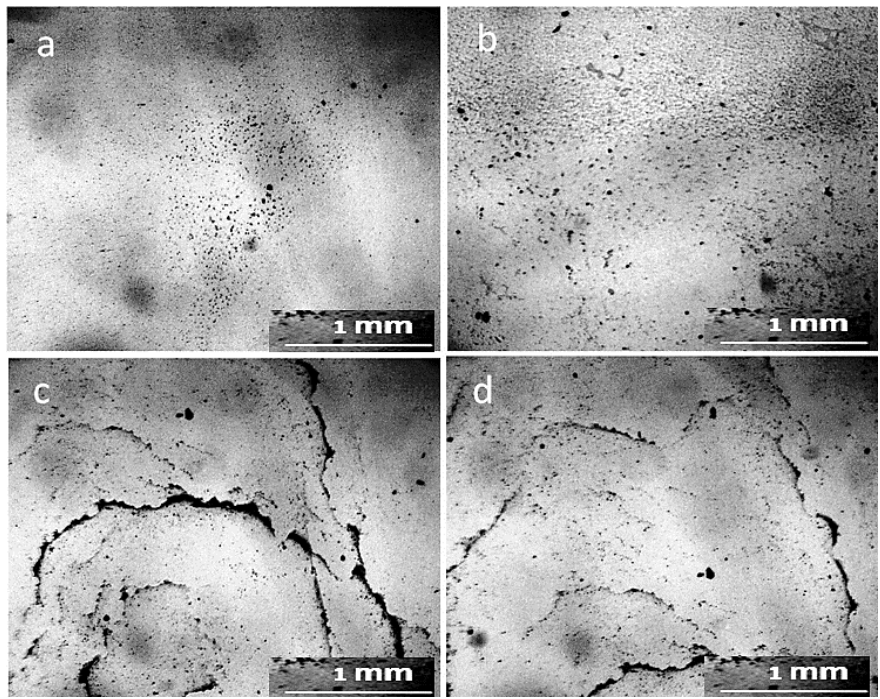

Figure 12. Microscopic Photograph. Middle part of incubated slides. a, b – deposition of DP aggregates in distilled water; c, d are layers of DP aggregates in tap water [43].

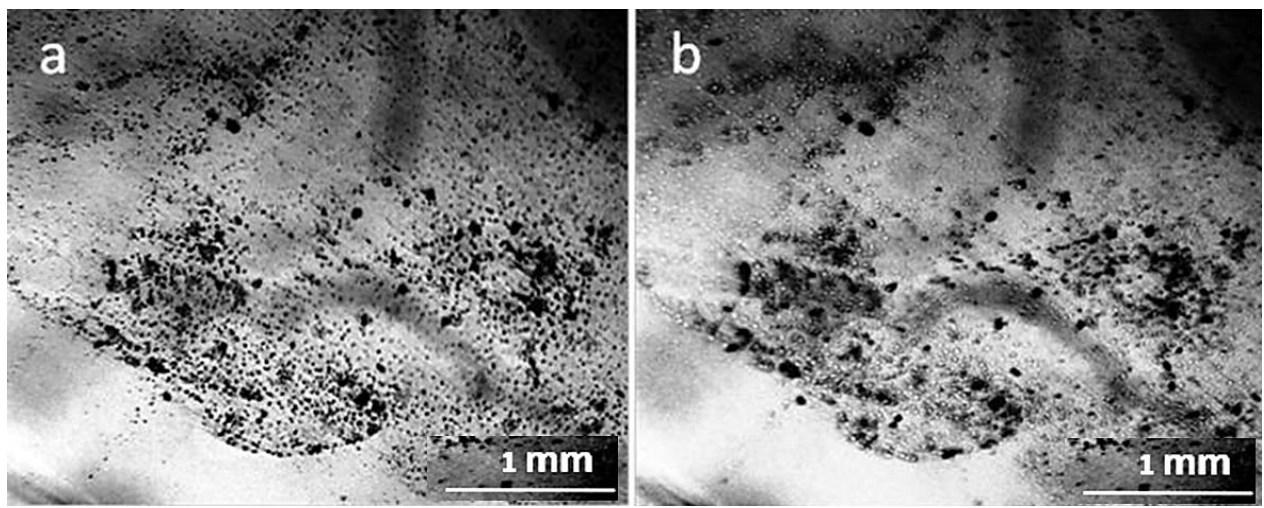

Figure 13. Fragment of the distilled water DP sediment at the bottom of the Petri dish: a – in normal viewing mode; b - with an increase in the lens depth of field. Salt microcrystals (white dots) are visible, the height of which exceeds the average height of the preparation due to the partial dissolution of hydration shells [43].

The experiments carried out confirm our ideas that NaCl microcrystals are protected from dissolution in bulk water by a liquid-crystal hydration shell, which breaks down at a temperature of ~ 300°C [26] and "melts" with an increase in osmotic pressure (Fig. 13). Experiments on the effect of the gliding arc plasma on water are indicative [44]. This transient type of discharge is non-equilibrium with a relatively high microarc temperature (about 1600–1800 ◦K). The treatment of water by the method described by the authors reduced its viscosity, pH, and surface tension, which, apparently, was a consequence of the melting of the DP hydration shells and the transition of polymeric water to the liquid phase. Exposure of Nanosecond Spark Discharge Treatment [45] to sitting water drops drastically changed crystallization in the droplets under plasma treatment proceeds in a different way than in identical dried droplets in the conventional mode. We believe that these changes are also associated with the destruction of the polymer structure of DP hydration shells under the action of high temperature.

2. What happens during distillation?

The quality of distilled water is directly related to the results of experiments. Therefore, the study of distillation processes is extremely important for an adequate evaluation of the results. The authors of [46] designed and manufactured a laboratory setup, which is an evaporator with a separator. The objects of study were evaporated solutions of a number of inorganic salts in the range of initial concentrations from 6 to 240 g/l. Salt solution of various initial concentrations was placed in the volume of the evaporator, after which the evaporation process took place and the condensate was taken. The resulting condensate was analyzed by the atomic adsorption spectroscopy method of analysis. The transition from free evaporation to regimes of unstable and stable boiling of the solution leads to a sharp increase in the entrainment of the dissolved substance from the evaporated solutions by secondary vapors. In the regimes of unstable and stable boiling, the main factor in the entrainment of a dissolved substance with steam is droplet entrainment [46].

We collected steam condensate over boiling tap water on a glass slide and examined it under a microscope. It turned out that the condensate dried on the glass contains microdroplets and fractal salt structures. Microdroplets, in turn, contain DP (Fig. 14).

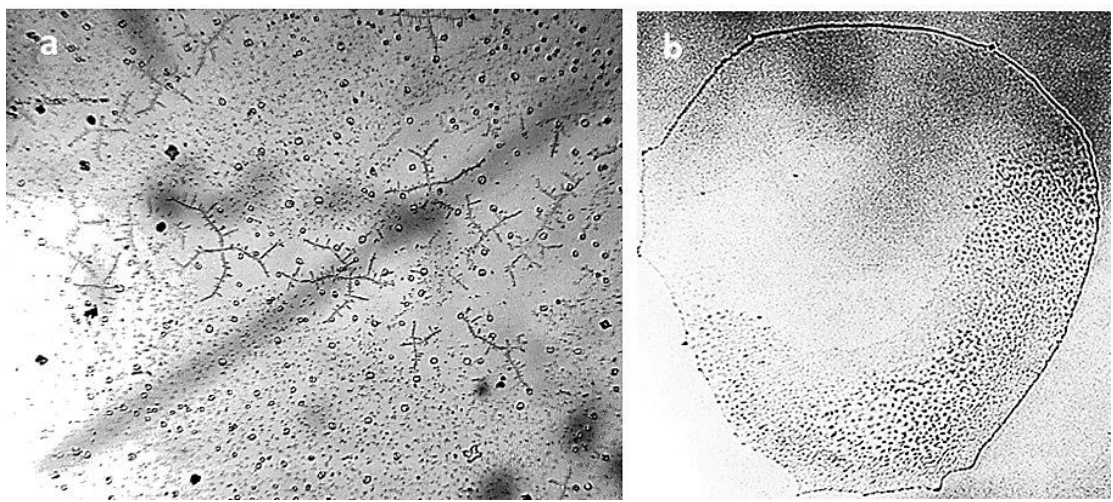

Figure 14. Condensate dried on glass (a) contains microdroplets and fractal salt structures. Frame width - 3 mm. The condensate microdroplet (b) contains a microdispersed phase. Frame width ~ 250 μm.

Thus, in a simple experiment, we have seen that both DP and salt are carried away with steam. According to the handbook [47], the solubility of NaCl in vapor is much higher than that of other salts. Therefore, we believe that the fractal salt clusters observed by us are formed by sodium chloride. We also observed the growth of similar fractal NaCl clusters in drying drops of albumin-saline solution [48] and in dried drops of ultra-high dilution saline solution with ultra-pure water ($10^{-7}$ M) [49; 25].

3. Conclusion

Within the framework of this review, we have tried to present the entire spectrum of the prevailing ideas about the supramolecular (on the scale of tens and hundreds of micrometers) structures of aqueous media. As can be seen from the information presented, there is no consensus among researchers. It remains to state the following.

1. Various methods of physical analysis have proved the presence in water, including a high degree of purification, of microparticles (clusters) ranging in size from units to hundreds of micrometers.

2. The authors of [10-14] believe that the Giant Heterophase Clusters are dynamic water associates with relaxation times from fractions to units of a second. According to our observations, the clusters are quite stable, and no "near-second relaxation times" are applicable to them. The disappearance of certain structures from the field of view is not associated with the disintegration of these structures, but with their spatial displacement as a result of Brownian motion.

3. According to our data, each DP microsphere contains a NaCl microcrystal inside. In this case, transitional forms are not observed. Therefore, the term "emulons" [15] is incorrect, since, in addition to a liquid of another phase, it also contains a particle of a solid substance.

4. The proposed nature of these particles, according to different authors, is different - from gas microbubbles (bubstons) to a dynamic ensemble of clusters of water molecules and NaCl microcrystals surrounded by a thick layer of hydrated liquid crystal water.

The latter version seems to us the most plausible, since it fits into the logic of the salt cycle in nature, the two-phase state of water in solutions [50], and SEM data. We have presented the results of our experiments and their logical reasoning to support our point of view. In the framework of this review, we did not touch upon quantum mechanical concepts of the structure and dynamics of aqueous media [51], which, undoubtedly, is an important part of future research